\documentclass[aps,prd,groupedaddress,showpacs,nofootinbib,amssymb,balancelastpage,preprintnumbers,floatfix]{revtex4}

\usepackage[dvipdfmx]{graphicx}
\usepackage{graphicx,bm,color}

\usepackage{amsmath}
\usepackage{amssymb}
\usepackage{amsfonts}
\usepackage{cases}
\usepackage{cancel}
\usepackage{hyperref}
\usepackage[normalem]{ulem}
\usepackage{subfigure}
\usepackage{here}
\usepackage{color}
\usepackage{comment}

\usepackage{tikz}
\usetikzlibrary{matrix}

\allowdisplaybreaks[3]

\begin{document}
	

\title{ 
	Gravitational wave footprints from  
	Higgs-portal scalegenesis \\
	with multiple dark chiral scalars 
}

\author{He-Xu Zhang}\thanks{{\tt hxzhang18@163.com}}
\affiliation{Center for Theoretical Physics and College of Physics, Jilin University, Changchun, 130012,
	China}

\author{Hiroyuki Ishida}\thanks{{\tt ishidah@pu-toyama.ac.jp}}
\affiliation{Center for Liberal Arts and Sciences, Toyama Prefectural University, Toyama 939-0398, Japan}

\author{Shinya Matsuzaki}\thanks{{\tt synya@jlu.edu.cn}}
\affiliation{Center for Theoretical Physics and College of Physics, Jilin University, Changchun, 130012,
	China}%

\begin{abstract}
We discuss the gravitational wave (GW) spectra predicted from 
the electroweak scalegenesis of the Higgs portal type 
with a large number of dark chiral flavors, which many flavor QCD would underlie and give the dynamical explanation of the negative Higgs portal coupling required to trigger the electroweak symmetry breaking. 
We employ the linear-sigma model as the low-energy description of dark many flavor QCD and show that the model undergoes ultra-supercooling due to the produced 
strong first-order thermal phase transition along the (approximately realized) flat direction based on the Gildener-Weinberg mechanism. 
Passing through evaluation of the bubble nucleation/percolation, we address 
the reheating and relaxation processes, which are generically non-thermal and nonadiabatic. 
Parametrizing the reheating epoch in terms of the e-folding number, we propose 
proper formulae for the redshift effects on the GW frequencies and signal 
spectra. 
It then turns out that the ultra-supercooling predicted from the Higgs-portal scalegenesis generically  
yields none of GW signals with the frequencies as low as nano Hz, unless the released latent heat is transported into another sector other than reheating the universe. 
Instead, models of this class prefer to give the higher frequency signals and 
still keeps the future prospected detection sensitivity, like at LISA, BBO, and DECIGO, etc. 
We also find that with large flavors in the dark sector, the GW signals are made further smaller 
and the peak frequencies higher. 
Characteristic phenomenological consequences related to 
the multiple chiral scalars include the prediction of dark pions 
with the mass much less than TeV scale, which is also briefly addressed. 
 
\end{abstract}

\maketitle

\section{Introduction}

The origin of mass and the electroweak symmetry breaking is not 
sufficiently accounted for in the standard model (SM), although 
the SM-like Higgs was discovered~\cite{Aad:2012tfa,Chatrchyan:2012xdj}: 
in the SM, the sign of the Higgs mass parameter is necessarily assumed to be negative, to realize the electroweak symmetry breaking, which is given by hand. 
This is indeed the longstanding and unsolved issue still left at present, with which the gauge hierarchy problem or fine tuning problem is also associated.

One idea to tackle this issue is to consider the so-called classical 
scale invariance. 
This is originated from the Bardeen's argument~\cite{Bardeen:1995kv}:   
the classical scale invariance sets the Higgs mass parameter to be zero at some scale in the renormalization group evolution, say, at 
the Planck scale, so that the Higgs mass will not be generated. 
It has been so far suggested that the classical scale invariance for the Higgs potential at the Planck scale 
can be realized as an infrared fixed point nonperturbatively 
generated by quantum gravitational effects~\cite{Shaposhnikov:2009pv,Wetterich:2016uxm,Eichhorn:2017als,Pawlowski:2018ixd,Wetterich:2019qzx}.

It might be interesting to argue also that the observed SM-like Higgs is supposed to have the profile along a nearly scale-invariant direction, i.e., the flat direction in the electroweak-broken phase. 
This can be manifested by realizing the fact that 
the small Higgs quartic coupling, 
$\lambda_H = (m_h^2/2v_h^2) \simeq 1/8 \ll 1$
with taking the limit $\lambda_H 
\to 0$ leads to the flat Higgs potential keeping 
nonzero $v_{h} \simeq 246$ GeV and 
the mass $m_h\simeq 125$ GeV~\cite{Matsuzaki:2016iyq,Yamawaki:2018jvy}.

Given the classical scale invariance, the scalegenesis has to be 
triggered by new physics, like a dark sector. 
The simplest idea along this conformal extension of the SM 
is to predict one SM-singlet scalar, $S$, 
allowing coupling to the Higgs doublet via forming the portal with  
a real scalar~\cite{Hempfling:1996ht}
or an extra $U(1)$-charged scalar~\cite{Meissner:2006zh}, 
or a generic complex scalar with or without $CP$ violation~\cite{AlexanderNunneley:2010nw,Farzinnia:2013pga,Gabrielli:2013hma}, 
such as $ |H|^2 S^2$. 
Those dark sector scalars together with 
the SM-like Higgs develop the flat direction and 
the classical scale invariance is spontaneously and 
explicitly broken by the dimensional transmutation at 
the quantum loop level, due to what is called 
Coleman-Weinberg mechanism~\cite{Coleman:1973jx} 
and/or Gildener-Weinberg mechanism~\cite{Gildener:1976ih}. 
This is the scalegenesis of one kind, what we may call the Higgs portal scalegenesis.

In the simplest Higgs-portal scalegenesis where  
only single singlet scalar is introduced, 
the portal coupling is necessarily assumed to be negative. 
There even including the radiative corrections, 
one needs to require the portal coupling to be negative by hand,  
otherwise any models can never realize the electroweak symmetry breaking (see, e.g., \cite{Sannino:2015wka}, 
and references therein). 
Actually, this is the same drawback as what the SM possesses 
in terms of the negative Higgs mass parameter. 
Therefore, the simple-minded 
Higgs portal scalegenesis still calls for some new physics.

One way out is to further predict an additional dark sector with a new gauge symmetry $U_{B-L}$ 
or $U(1)_X$. 
In that case the new scalar is charged under the new gauge symmetry 
(e.g. the $B-L$ Higgs), so that the negative portal coupling can be generated at low energy by 
 the renormalization group evolution, as has been addressed in the literature~\cite{Iso:2012jn,Hashimoto:2014ela,Hamada:2020vnf}.

Another type of the dynamical origin of 
the negative portal coupling has been proposed 
in a unified way in~\cite{Ishida:2019gri}. 
It is mandatory to link with an underlying (almost) 
scale-invariant dark QCD with many flavors. 
In this scenario, the Higgs portal partner, a dilaton, 
arises as a composite-singlet scalar generated from the underlying scale-invariant many flavor gauge theory. 
The scale anomaly induced 
via the Gildener-Weinberg/Coleman-Weinberg mechanism 
can also be interpreted, by the anomaly matching, as the nonperturbative scale anomaly coupled to the composite dilaton, where the latter is generated by the dynamical chiral-scale breaking in 
the underlying theory. 
Along this scenario, generically plenty of dark hadron spectra will be predicted due to the many flavor structure, which could be testable at collider experiments and/or through footprints left in cosmological observations. 
Such an almost-scale invariant feature has also been applied to inflationary scenarios 
with the small-field inflation of 
the Coleman-Weinberg  type~\cite{Ishida:2019wkd,Zhang:2023acu}.

In this paper, we focus on many flavor QCD scenario in a view of 
the underlying theory for the Higgs portal scalegenesis, 
and discuss the gravitational-wave (GW) footprints in cosmology arising from 
the cosmological phase transition along the flat direction. 
We in particular take the number of dark flavors ($N_f$) to be $8$ for scale-invariant 
many flavor QCD with the number of colors $N_c=3$, as a definite benchmark model, though we will keep 
arbitrary $N_f$ when discussing analytic features. 
This setup has been definitely clarified, in lattice simulations, to be scale-invariant QCD along with 
presence of the chiral broken phase~\cite{LSD:2014nmn,Aoki:2014oha,Hasenfratz:2014rna} and the light composite dilaton~\cite{Aoki:2016wnc,LatKMI:2016xxi,Appelquist:2016viq,LatticeStrongDynamics:2018hun} 
(when the eight fermions are in the fundamental representation of the gauge group).

We work on the scale-invariant linear sigma model as 
the low-energy description of underlying many flavor QCD, to which  
the SM sector couples through the Higgs portal. 
With the currently available observables and constraints related to the Higgs sector at hand,  
we analyze the cosmological phase transition and show that in the case of the benchmark model with 
$N_f=8$ the ultra-supercooling is generated and the nucleation/percolation of true-vacuum bubbles. The large flavor dependence on the cosmological phase transition is also discussed.  Then we evaluate the GW signals sourced from the ultra-supercooling.

In the literature~\cite{Miura:2018dsy}, GW spectra produced from the ultra-supercooling in  
many flavor QCD with $N_f=8$ have been discussed based on the scale-invariant linear sigma model description as the low-energy effective theory. 
This is, however, not the Higgs portal scalegenesis, 
but what is called (many-flavor) walking technicolor~\cite{Yamawaki:1985zg}, where the composite dilaton (called  
technidilaton~\cite{Bando:1986bg}) plays the role of the SM-like Higgs itself.

In other works~\cite{Salvio:2023qgb,Salvio:2023ynn} 
the ultra-supercooling generated from the Higgs portal scalegenesis coupled to 
multiple SM singlet scalars have been discussed in a generic manner in light of prediction of 
the GW signals. 
In the present study, the SM-like Higgs forms the portal coupling 
only to a singlet scalar (dilaton), not multiple of scalars like in the literature. 
The flat direction, derived from the present scenario, is thus the simplest, in contrast to 
the one in the literature.


Our particular claim is also on evaluation of the redshift effect on 
the produced GWs. 
This redshift arises through the reheating epoch due to releasing the false vacuum energy (latent heat) into the SM thermal plasma via the Higgs portal coupling. 
Parametrizing the reheating epoch in terms of the e-folding number, we propose 
proper formulae for the redshift effects on the GW frequencies and signal 
spectra.

We find that the ultra-supercooling with large $N_f$ generically  
yields none of GW signals with the frequencies as low as nano Hz (namely, no signal in NANO Grav 15yr~\cite{NANOGrav:2023gor} and also in other nano Hz signal prospects~\cite{Xu:2023wog,EPTA:2023fyk,Reardon:2023gzh}), instead, prefers to give the higher frequency signals.  
The thus characteristically produced GW spectra, however, still keep having the future prospected detection sensitivity, like at the Laser Interferometer Space Antenna (LISA)~\cite{LISA:2017pwj,Caprini:2019egz}, 
the Big Bang Observer (BBO)~\cite{Corbin:2005ny,Harry:2006fi}, and Deci-hertz Interferometer Gravitational Wave Observatory (DECIGO)~\cite{Kawamura:2006up,Yagi:2011wg}, etc. 
We also find that with large $N_f$, the GW signals are made further smaller 
and the peak frequencies higher.

This paper is organized as follows: 
in Sec.~II the model for the Higgs portal scalegenesis with multiple dark chiral scalars is introduced in details 
and the flat direction at the tree-level as well as the phenomenological constraint 
to fix the model parameters are discussed. 
In Sec.~III we show the one-loop computations of the effective potential arising 
from the model introduced in Sec.~II, based on the Gildener-Weinberg mechanism 
and the standard way of incorporation of thermal corrections. 
In Sec.~IV the cosmological-first order-phase transition predicted from the present model 
is addressed in details, including the ultra-supercooling phenomenon and 
the nucleation of the created bubbles. There we find the characteristic features for 
the phase transition parameters (denoted as $\alpha$ and $\beta$) closely 
tiled with the consequence of the ultra-supercooling in the Higgs-portal scalegenesis. 
Section V provides the evaluation of the GWs sourced from 
the predicted ultra-supercooling in details. Then we propose a proper formula to take into account 
the redshift effect on the produced GWs related to reheating of the universe. 
Summary of the present study is given in Sec.~V, where we also discuss phenomenological 
and cosmological consequences, other than the predicted GW signals, 
including cosmology of possible dark matter candidates and collider experimental probes 
for the dark sector with many chiral flavors, such as the prediction of dark pions 
with the mass much less than TeV scale.

\section{The model set-up}
In this section, we begin by modeling the Lagrangian having the  chiral $U(N_f)_L \times U(N_f)_R$ symmetry and the classical scale invariance at some ultraviolet scale (above TeV). 
The building blocks consist of the so-called chiral field $M(x)$, 
which forms an $N_f \times N_f$ matrix and transforms under the global chiral 
$U(N_f)_L \times U(N_f)_R$ symmetry as
\begin{equation}
	M(x) \rightarrow g_L \cdot M(x) \cdot g^\dagger_R,\quad g_L,g_R \in U(N_f)\,, 
\end{equation}
and its hermitian conjugate $M^\dag$, 
where $g_L$ and $g_R$ stand for the transformation matrices belonging  
to the chiral-product group $U(N_f)_L \times U(N_f)_R$. 
This $M$ transforms under the scale (dilatation) symmetry to get the infinitesimal shift as 
\begin{align}
    \delta_D M(x) = (1 + x^\nu \partial_\nu) \cdot M(x) 
    \,.   
\end{align}
In addition, we impose the parity ($P$) and charge conjugate ($C$) 
invariance in the $M$ sector, which transform $M$ as 
$M \to M^\dag$ for $P$, and $M \to M^T$ for $C$. 
Hereafter we will suppress the spacetime coordinate dependence on fields, unless necessary. 

Including the SM-like Higgs field coupled to this $M$ 
in a manner invariant under the global chiral 
$U(N_f)_L \times U(N_f)_R$ and SM gauge symmetries 
together with $C$ an $P$ invariance in the $M$ sector, 
we thus construct the scale-invariant linear sigma model with 
the scale-invariant SM as follows: 
\begin{equation}
	\mathcal{L} = \mathcal{L}_{\rm \overline{SM}}\, + \mathrm{Tr} \left[\partial_\mu M^\dagger\partial^\mu M\right] - V(H,M)\,,
\end{equation}
where $\mathcal{L}_{\rm \overline{SM}}$ is the SM Lagrangian without the Higgs potential term, 
and $V(H,M)$ denotes the scale-invariant potential which takes the form 
\begin{equation}
	\label{potential}
	V(H,M) = \lambda_1 \left(\mathrm{Tr}[M^\dagger M]\right)^2 + \lambda_2 \mathrm{Tr}\left[(M^\dagger M)^2\right] + \lambda_{\rm mix} |H|^2 \mathrm{Tr}[M^\dagger M]+\lambda_h |H|^4
\,, 
\end{equation} 
with $\lambda_h$, $\lambda_1$, 
and $\lambda_2$ being positive definite, while $\lambda_{\rm mix}$ negative.

The chiral $U(N_f)_L \times U(N_f)_R$ symmetry is assumed to be spontaneously broken down to the diagonal subgroup $U(N_f)_V$, what we shall conventionally call the dark isospin symmetry. 
The $U(N_f)_V$ symmetry is manifestly unbroken reflecting the underlying QCD-like theory as the vectorlike gauge theory. 
Taking into account this symmetry breaking pattern 
and the VEV of the SM Higgs field $H$ 
to break the electroweak symmetry as well, $M$ and $H$ fields are parametrized as 
\begin{equation}
	\langle M \rangle =\frac{\phi}{\sqrt{2 N_f} }\cdot \,\mathbb{I}_{N_f \times N_f} \,,\quad \langle H \rangle=\frac{1}{\sqrt{2}}\binom{0}{h}\,, 
\end{equation}
where $\mathbb{I}_{N_f \times N_f}$ is the $N_f$ by $N_f$ unit matrix. 
In terms of the background fields $\phi$ and $h$, 
the tree-level potential is thus read off from Eq.(\ref{potential}) as 
\begin{equation}
	\label{treelevelV}
	V_{\rm tree} = \frac{1}{4} \left( \lambda_1 + \frac{\lambda_2}{N_f} \right)\phi^4 + \frac{\lambda_{\rm mix}}{4} h^2 \phi^2 + \frac{\lambda_h}{4} h^4\,.
\end{equation}
To this potential, the potential stability condition requires   
\begin{equation}
	\lambda_h \ge 0, \qquad \lambda_{\rm mix}^2\le 4\left(\lambda_1+\frac{\lambda_2}{N_f}\right)\lambda_h\,. 
\end{equation}

We apply the Gildener-Weinberg approach~\cite{Gildener:1976ih} 
and try to find the flat direction, which can be oriented along 
$h \propto \chi$ and $\phi \propto \chi$ with 
the unified single background $\chi$. 
This proportionality to $\chi$ is clarified by solving mixing between $h$ and $\phi$ arising 
in $V_{\rm tree}$ of Eq.(\ref{treelevelV}) with the mixing angle $\theta$, in such a way that  
\begin{align}
h&= \chi \sin\theta \,, \notag\\ 
\phi& = \chi\cos\theta    
\,. \label{h-chi}
\end{align} 
Using this we rewrite the tree-level potential in Eq.(\ref{treelevelV}) 
as a function of $\chi$, to get 
\begin{equation}
	V_{\rm tree} = \frac{\chi^4}{4}\left[ \left( \lambda_1 + \frac{\lambda_2}{N_f} \right)\cos^4\theta + \lambda_{\rm mix} \cos^2\theta\sin^2\theta + \lambda_h \sin^4\theta\right]. 
\end{equation}
The flat direction condition, which requires $V_{\rm tree}$ to vanish and stationary along that direction, yields
\begin{equation}
	\tan^2\theta = \frac{-\lambda_{\rm mix}}{2\lambda_h}\,, \quad \quad 
	\lambda_{\rm mix}^2 = 4 \left( \lambda_1 + \frac{\lambda_2}{N_f} \right)\lambda_h\, , 
\label{FD}
\end{equation}
at certain renormalization group scale $\mu$.

Around the VEV in the flat direction $M$ and $H$ can be expanded as
\begin{equation}
	\label{matrixfield}
	M=\frac{\phi+\sigma+i\eta}{\sqrt{2N_f}}\cdot \mathbb{I}_{N_f \times N_f}+\sum_{a=1}^{N_{f}^{2}-1}\left(\xi^a+i \pi^{a}\right) T^{a}, \quad  H = \frac{1}{\sqrt{2}} \binom{0}{h+\tilde{h}}\,,
\end{equation}
where $T_a$ are the generators of $SU(N_f)$ group in the fundamental representation and normalized as
${\rm Tr}(T^aT^b)=\delta^{ab}/2$, and $\tilde{h}$ denotes the Higgs fluctuation field.   
In Eq.(\ref{matrixfield}) $\sigma$ and $\eta$ are the dark isospin-singlet scalar and 
pseudoscalar fields, while $\xi^a$ and $\pi^a$ the dark isospin-adjoint scalar and pseudoscalar fields, respectively.  
These dark-sector fields would be regarded as mesons in terms of the expected underlying QCD-like gauge theory.  
The field-dependent mass-squares for $\tilde{h}$, $\sigma$, $\eta$, $\xi^a$, and $\pi^a$ 
then read  
\begin{align}
m_{\sigma}^2(\chi) 
&= 3 \left( \lambda_1+\frac{\lambda_2}{N_f} \right)\chi^2\cos^2\theta + \frac{\lambda_{\rm mix}}{2} \chi^2 \sin^2\theta = 2 \left( \lambda_1+\frac{\lambda_2}{N_f} \right)\chi^2\cos^2\theta\,, \notag \\
m_{\xi^a}^2(\chi) &= \left(\lambda_1 +\frac{3\lambda_2}{N_f}\right)\chi^2\cos^2\theta + \frac{\lambda_{\rm mix}}{2} \chi^2 \sin^2\theta =\frac{2\lambda_2}{N_f}\chi^2\cos^2\theta\,, \notag \\
m_\eta(\chi) &= m_{\pi^a}^2(\chi) = 0\,, \notag\\ 
m_{\tilde{h}}^2(\chi) &= -\lambda_{\rm mix} \chi^2\cos^2\theta\,,  
\end{align}
where we have used the flat direction condition in Eq.(\ref{FD}). 
Note that the Nambu-Goldstone bosons $\pi^a$ and $\eta$ 
associated with the spontaneous chiral breaking are surely massless along the flat direction. 

The angle $\theta$ defined in Eq.(\ref{h-chi}) simultaneously diagonalizes 
the $h\mathchar`-\chi$ mixing mass matrix, 
\begin{equation}
	\mathcal{M}^2=\begin{pmatrix}
		2\lambda_h v_h^2 &  \lambda_{\rm mix}v_h v_\phi\\
		\lambda_{\rm mix}v_h v_{\phi} &\displaystyle 2 \left(\lambda_1+\frac{\lambda_2}{N_f} \right) v_{\phi}^2
	\end{pmatrix}\,, 
\end{equation}
in such a way that 
\begin{equation}
	\binom{h_1}{h_2}= \begin{pmatrix}
		\cos\theta &  -\sin\theta\\
		\sin\theta & \cos\theta
	\end{pmatrix} \binom{\tilde{h}}{\sigma}\,,  
\end{equation}
with the mass eigenstate fields $h_1$ and $h_2$. 
This eigenvalue system gives the tree level mass eigenvalues for 
the mass eigenstates $h_1$ and $h_2$ as  
\begin{equation}
	m_{h_1}^2=-\lambda_{\rm mix}v_\chi^2 \quad , \quad m_{h_2}^2=0\,.
\label{lambda-mix}
\end{equation}
At this moment, $h_2$ thus becomes massless (called the scalon~\cite{Gildener:1976ih}) having the profile along the flat direction. At the one-loop level, this $h_2$ acquires a mass as the flat direction is lifted by the quantum corrections, and becomes what is called the pseudo-dilaton due to the radiative scale symmetry breaking. On the other hand, $h_1$ has the profile  perpendicular to the flat direction, identified as the SM-like Higgs, observed at the LHC with $m_{h_1} \simeq  125$ GeV, which does not develop its mass along the flat direction.

Current experimental limits 
on the mixing angle $\theta$ can be read off from the total signal strength of the Higgs 
coupling measurements at the Large Hadron Collider (LHC)~\cite{ParticleDataGroup:2022pth}. 
The limit can conservatively be placed as 
\begin{equation}
    \sin^2\theta  = \frac{v_h^2}{v_\chi^2}\lesssim 0.1\,,\quad \text{i.e.,}\quad v_\chi \gtrsim 778\,\text{GeV}, 
\label{Higgs-coupling}
\end{equation}
with $v_h \simeq 246$ GeV being fixed to the electroweak scale.

\section{Gildener-Weinberg type scalegenesis and thermal corrections}

In this section, along the flat direction in Eq.(\ref{FD}), 
we compute the one-loop effective potential at zero temperature in the $\overline{\rm MS}$ scheme~\footnote{We take the Landau gauge for the SM gauge loop contributions. In this case the Nambu-Goldstone boson loop contributions are field-independent at the leading order in the resummed perturbation theory, so that they are decoupled in the effective potential analysis.}. 
The thermal corrections are then incorporated in an appropriate way at the consistent one-loop level.

We find that the resultant one-loop effective potential at zero temperature takes the form   
\begin{equation}
	V_1(\chi) = A \chi^4 +B \chi^4 \log \frac{\chi^2}{\mu_{\rm GW}^2}\,,
\end{equation}
with
\begin{align}
	A =& \frac{\cos^4\theta}{16\pi^2}\Bigg[\left( \lambda_1+\frac{\lambda_2}{N_f} \right)^2 \left(\log \left( 2 \left( \lambda_1+\frac{\lambda_2}{N_f} \right) \cos^2\theta\right)-\frac{3}{2}\right) + (N_f^2-1)\frac{\lambda_2^2}{N_f^2}\left(\log \left( \frac{2\lambda_2}{N_f}\cos^2\theta\right)-\frac{3}{2}\right) 
 \notag \\
	&+\frac{\lambda_{\rm mix}^2}{4}\left(\log \left(|\lambda_{\rm mix}|\cos^2\theta\right)-\frac{3}{2}\right)\Bigg]
	+\frac{1}{64\pi^2 v_\chi^4}\sum_{i=t,Z,W^\pm} (-1)^{s}n_i m_i^4\left(\log \frac{m_i^2}{v_\chi^2}-c_i\right), 
 \notag \\
	B =&\frac{\cos^4\theta}{16\pi^2}\left[ \left( \lambda_1+\frac{\lambda_2}{N_f} \right)^2 + \left( N_f^2-1 \right)\frac{\lambda_2^2}{N_f^2}+\frac{\lambda_{\rm mix}^2}{4}\right] + \frac{1}{64\pi^2 v_\chi^4}\sum_{i=t,Z,W^\pm} (-1)^{s}n_i m_i^4\,,
\label{A-B}
\end{align}
where $s=1\, (0)$ for fermions (bosons); $c_i=\frac{1}{2}\,(\frac{3}{2})$ for the transverse (longitudinal) polarization of gauge bosons, and $c_i=\frac{3}{2}$ for the other particles. The numbers of degree of freedom (d.o.f.) $n_i$ for $i = t$, $Z$, $W^\pm$ are 12, 3, 6, respectively, and their masses can be written as $m_i^2(\chi)=m_i^2\frac{\chi^2}{v_\chi^2}$. 

The nonzero VEV of $\chi$ is associated with the renormalization scale $\mu_{\rm GW}$ via the stationary condition (as the consequence of the dimensional transmutation):
\begin{equation}
	\frac{\partial V_1(\chi)}{\partial \chi}=0 \quad \Rightarrow \quad \mu_{\rm GW} = v_\chi \exp \left( \frac{A}{2B}+\frac{1}{4} \right)\,.
\end{equation}
Correspondingly, the effective potential can be rewritten as
\begin{equation}
	V_1(\chi) = B \chi^4 \left(\log \frac{\chi^2}{v_\chi^2}-\frac{1}{2}\right)+V_0\,,\quad V_0=\frac{B v_\chi^4}{2} 
 \simeq \frac{\lambda_2^2 v_\chi^4}{32 \pi^2} \frac{N_f^2-1}{N_f^2}
 \,, \label{V1}
\end{equation}
from which 
the radiatively generated mass of $\chi$ is also obtained as 
\begin{equation}
	M_\chi^2= \left.\frac{\partial^2 V_1(\chi)}{\partial \chi^2}\right|_{\chi = v_\chi} = 8 B v_\chi^2\,.
\label{Mchi}
\end{equation} 
In Eq.(\ref{V1}) $V_0$ denotes the vacuum energy, which is determined by the normalization condition $V_1(v_\chi)=0$, and the last approximation has been made by taking into account the flat direction condition 
Eq.(\ref{FD}) together with the constraints from realization of the Higgs mass and the electroweak scale in Eqs.(\ref{Higgs-coupling}) and (\ref{lambda-mix}). 
Note that the potential stability condition $B>0$ at one-loop level is trivially met in the present model. 


To be phenomenologically realistic, 
we need to introduce an explicit scale and chiral breaking term, otherwise 
there are plenty of massless Nambu-Goldstone bosons, $\pi^a$ and $\eta$, 
left in the universe. 
However, 
as long as the explicit breaking small enough that 
the flat direction can still approximately work, the 
to-be-addressed characteristic features on the cosmological phase transition 
and the gravitational wave production will not substantially be altered. 
Later we will come back to this point in a view of the phenomenological 
consequences related to the predicted GW spectra (see Summary and Discussion).

By following the standard procedure, 
the one-loop thermal corrections are evaluated as 
\begin{align}
	V_{1,T}(\chi,T) = &\frac{T^4 }{2\pi^2} J_B\left(\frac{m_\sigma^2(\chi)}{T^2}\right) +  \frac{\left(N_f^2-1\right)T^4 }{2\pi^2} J_B\left(\frac{m_{\xi^i}^2(\chi)}{T^2}\right) + \frac{T^4 }{2\pi^2} J_B\left(\frac{m_h^2(\chi)}{T^2}\right) \notag \\
	 & +\frac{T^4 }{2\pi^2} \left[\sum_{i=t,Z,W} (-1)^{2s} n_i J_{B/F}\left(\frac{m_i^2(\chi)}{T^2}\right)\right]\,, \label{V1T}
\end{align}
with the bosonic/fermionic thermal loop functions
\begin{equation}
	J_{B/F}(y^2) = \int_{0}^{\infty}\mathrm{d}t\, t^2\ln\left(1\mp e^{-\sqrt{t^2+y^2}}\right)\,.
\end{equation}
It has been shown that the perturbative expansion will break down since in the high-temperature limit higher loop contributions can grow as large as the tree-level and one-loop terms~\cite{Curtin:2016urg,Senaha:2020mop}. 
To improve the validity of the perturbation, we adopt the \textit{truncated full dressing} resummation procedure~\cite{Curtin:2016urg}, which is performed by the replacement $m_{i}^2(\chi)\rightarrow m_{i}^2(\chi) +\Pi_i(T)$ in the full effective potential. 
The thermal masses $\Pi_i(T)$ are computed as follows: 
\begin{align}
	&\Pi_{\sigma/\xi^i}(T) = \frac{T^2}{6}\left[ \left( N_f^2+1 \right) \lambda_1+2N_f\lambda_2+\frac{\lambda_{\rm mix}}{4}\right],\quad
	\Pi_h(T) = T^2\left(\frac{\lambda_h}{4}+\frac{y_t^2}{4}+\frac{3g^2}{16}+\frac{{g'}^2}{16}+\frac{\lambda_{\rm mix}}{24}+\frac{N_f^2}{12}\lambda_{\rm mix}\right), \notag\\
	&\Pi_W^L(T) = \frac{11}{6}g^2T^2,\quad \Pi_W^T(T) = 0,\quad \Pi_Z^L(T) = \frac{11}{6}(g^2+{g'}^2)T^2,\quad \Pi_Z^T(T) = 0\,.
\end{align}
Here the SM gauge and Yukawa couplings are defined through 
the masses of $W$, $Z$ bosons, and top quark, respectively, 
as $g=2m_W/v_h$, $g'=2\sqrt{m_Z^2-m_W^2}/v_h$, and $y_t=\sqrt{2}m_t/v_h$. 
$\Pi_{W/Z}^L(T)$ denotes the thermal masses of the longitudinal mode of $W$ or $Z$ bosons, while transverse modes $\Pi_{W/Z}^T(T)$ are protected not to be generated due to the gauge invariance. 
\textcolor{black}{In general, the contributions from the daisy resummation are less important due to the fact that the phase transition completes well below the critical temperature in the supercooling case. However, the thermal mass with such large number of degrees of freedom, $\mathcal{O}(N_f^2)$, will become ten times as big as the field-dependent mass around the barrier, so that it's necessary to include the daisy contributions.}

Taking into account the flat direction condition in Eq.(\ref{FD}) together with the inputs for 
the Higgs mass $m_h$, the electroweak scale $v_h$, and the SM gauge and top quark masses, 
we see that the total one-loop effective potential, $V_1$ in Eq.(\ref{V1}) plus $V_{1,T}$ in Eq.(\ref{V1T}), is evaluated as a function of $\lambda_2$ and $v_\chi$. 
From the next section, we shall discuss the cosmological phase transition 
in this parameter space.

\section{Cosmological phase transition: ultra-supercooling and nucleation}


In this section we address the cosmological phase transition based on the 
one-loop effective potential derived in the previous section. 
Since it is of the Coleman-Weinberg type, 
the phase transition becomes of first order to be strong enough, i.e., $\chi /T_{c} \gg 1$, in a wide range of the coupling parameter space, 
where $T_c$ denotes the critical temperature at which the false and true vacua get degenerated.

\begin{figure}[t]
	\includegraphics[scale=0.7]{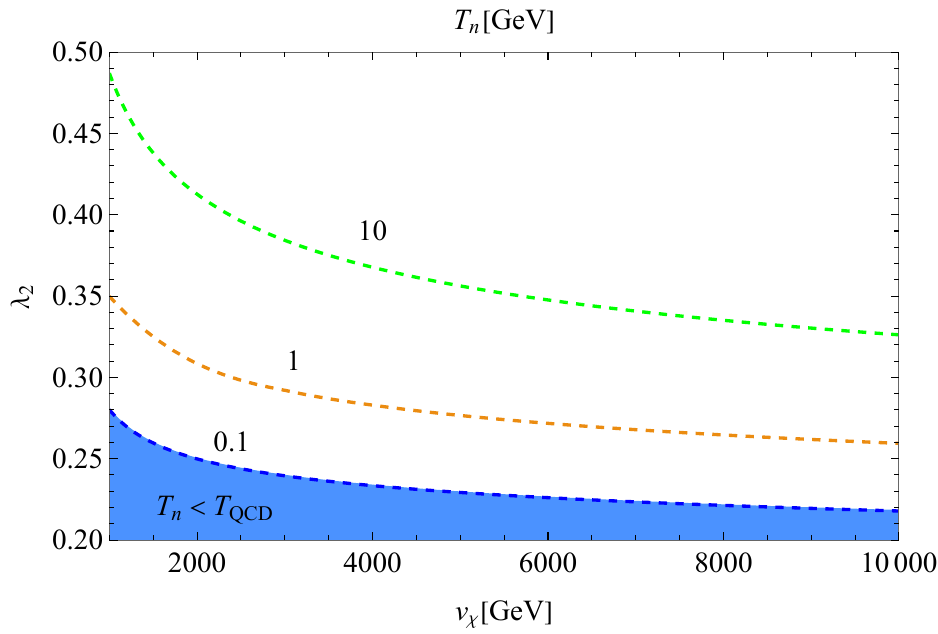}
	\caption{\label{fig:nucleation} 
		The contour plot of the nucleation temperature $T_n$ in the $(\lambda_2, v_\chi)$ plane 
  with $N_f=8$. 
  The blue-shaded regime corresponding to the case with $T_n < T_{\rm QCD}$ is discarded in the present study, which will actually be covered with the null percolation regime due to too small size of $\lambda_2$. } 
\end{figure}

In the expanding universe the first order phase transition proceeds by the bubble nucleation.  
The nucleation rate per unit volume/time of the bubble, $\Gamma(T)$, can be computed as  
\begin{equation}
	\Gamma(T) \simeq T^4\left(-\frac{S_3(T)}{2\pi T}\right)^{3/2}\exp\left(-\frac{S_3(T)}{T}\right),
\end{equation}
where $S_3(T)$ is the $\mathcal{O}(3)$ symmetric bounce action at $T$: 
\begin{equation}
	S_3(T) = 4\pi\int_0^\infty \mathrm{d}^3r\, r^2\left(\frac{1}{2}\left(\frac{\mathrm{d}\bar{\chi}}{\mathrm{d}r}\right)^2+V_{\rm eff}(\bar{\chi},T)\right).
\end{equation}
The normalizable bubble profile $\bar{\chi}(r)$ can be obtained by numerically solving the equation of motion,  
\begin{equation}
	\frac{\mathrm{d^2}\bar{\chi}}{\mathrm{d}r^2}+\frac{2}{r}\frac{\mathrm{d}\bar{\chi}}{\mathrm{d}r}=\frac{\mathrm{d}V_{\rm eff}(\bar{\chi},T)}{\mathrm{d}\bar{\chi}}
 \,, 
\end{equation}
with the boundary conditions
\begin{align}
    \left.\frac{2}{r}\frac{\mathrm{d}\bar{\chi(r)}}{\mathrm{d}r}\right|_{r=0}=0,\quad \left.\bar{\chi}(r)\right|_{r=\infty}=0\,. 
\end{align} 
The nucleation temperature $T_n$ is defined 
when the bubble nucleation rate for the first time catches up with the Hubble expansion rate:   
\begin{equation}
	\frac{\Gamma(T_n)}{H(T_n)^4} \sim 1\,,
\end{equation}
namely,
\begin{equation}
	\frac{S_3(T_n)}{T_n}-\frac{3}{2}\log\left(\frac{S_3(T_n)}{2\pi T_n}\right) 
 \sim 4\log\frac{T_n}{H(T_n)},
\end{equation}
where $H^2(T)=\left[\Delta V(T)+\rho_{\rm rad}(T)\right]/3M_{\rm pl}^2$, which, 
for the supercooled phase transition, can be well approximated by 
the vacuum energy part $H_V=\Delta V(T_n)/3M_{\rm pl}^2$.  
In Fig.~\ref{fig:nucleation} we display the contour plot of $T_n$ in 
the parameter space on the $(\lambda_2, v_\chi)$ plane for a couple of reference values for $T_n$ up to 10 GeV. 
There we have taken $N_f = 8$ as a benchmark 
inspired by underlying many flavor QCD as noted in the Introduction. 
In the plot we have discarded the case with $T_n < T_{\rm QCD}$ 
 because in that case instead of the Higgs portal, the QCD phase transition would trigger the electroweak phase transition, as addressed in 
the literature~\cite{Witten:1980ez,Iso:2017uuu,Hambye:2018qjv,Sagunski:2023ynd}, which is to be 
beyond our current scope.

The contour plot shown in Fig.~\ref{fig:nucleation} is qualitatively identical to the one discussed in~\cite{Kawana:2022fum} except for the size of the relevant couplings.  
The discrepancy comes from the quite different number of the dark-sector particles contributing to the one-loop effective potential: the present case is, say, $N_f^2$ (see Eqs.(\ref{V1}) and (\ref{V1T})), while the model in the literature only includes one. 
In particular since a large number of thermal loop contributions are created in the present model, 
the smaller size of the coupling is sufficient to achieve the nucleation over the Hubble rate. 
The percolation will process qualitatively in a similar manner as well. 
In the literature, it has been shown that due to too small size of the coupling strength, 
the null percolation regime is fully overlapped with 
the region for $T_n < T_{\rm QCD}$, which starts when $v_\chi$ gets as large as $\sim 10^4$ GeV, 
The percolation temperature $T_p$ has also been clarified to be almost identical to $T_n$ in a wide parameter space as in the contour plot, Fig.~\ref{fig:nucleation}. 
These features follow also in the present model.


The GW spectrum resulting from the cosmological-first order phase transition 
can be parametrized by two parameters $\alpha$ and $\beta$. 
The former $\alpha$ measures the strength of the first order phase transition, which is given by
the ratio of the latent heat released from the false vacuum to the radiation energy density: 
\begin{equation}
	\alpha \equiv \frac{1}{\rho_{\rm rad}(T_n)}\left(-\Delta V(T_n)+T_n\left.\frac{\mathrm{d}\Delta V(T)}{\mathrm{d}T}\right|_{T=T_n}\right)\simeq \frac{\Delta V(T_n)}{\rho_{\rm rad}(T_n)},
\label{alpha}
\end{equation}
where $\Delta V(T)$ is the difference of the effective potential at the true and false vacua~\footnote{
Instead of the latent, one can also use the trace of the energy-momentum tensor to define $\alpha$ (see, e.g.,~\cite{Athron:2023rfq,Athron:2023xlk}). However, they are equivalent each other in the case of the ultra-supercooled phase transition, 
because $\Delta V(T_n)\gg T_n \mathrm{d}\Delta V(T_n)/\mathrm{dT}$.}. 
The value of $\alpha$ turns out to be extremely large, $\alpha\gg 1$, for the ultra-supercooled phase transition.
The latter parameter $\beta$ and its normalized one $\tilde{\beta}$ are defined as 
\begin{equation}
	\tilde{\beta}\equiv\frac{\beta}{H(T_n)} = T_n\left.\frac{\mathrm{d}}{\mathrm{d}T}\left(\frac{S_3(T)}{T}\right)\right|_{T=T_n}\,,
\end{equation}
which measures the duration of the phase transition and the characteristic frequency of the GW through the mean bubble radius at collisions.

Another remark should be made on the characteristic temperature directly related to the peak frequencies of GWs, that is the reheating temperature $T_{r}$. 
It is usually argued that the estimate of $T_{r}$ depends on whether the rate of the $\chi$ decay to the SM sector ($\Gamma_{\rm dec}$) becomes smaller or larger than the Hubble parameter, that we shall classify in more details below. 

\begin{itemize} 

\item[(i)] 
In the case with $\Gamma_{\rm dec} \gg H(T_p)$, where the reheating is supposed to be processed instantaneously after the end of the supercooling, and the whole energy accumulated at the false vacuum is expected to be immediately converted into the radiation. 
The resultant reheating temperature is determined by assuming the full conversion of the vacuum energy into the radiation~\cite{Hambye:2018qjv,Ellis:2018mja} 
\begin{align}
	\rho_{\rm rad}(T_{r}) \simeq \rho_{\rm rad}(T_p) & + \rho_{\rm vac}(T_p) \simeq \rho_{\rm vac}(T_p) \notag \\
	\Rightarrow \qquad 
 T_{r} \simeq (1+\alpha)^{1/4} T_p &\simeq  \left(\frac{30\Delta V}{\pi^2 g_{r}}\right)^{1/4} \equiv T_{\rm vac}\,,  
\label{Treh-alpha}
\end{align}
where in the last line we have taken into account $\alpha \gg 1$ for the ultra-supercooling case.

\item[(ii)] 
In the case with $\Gamma_{\rm dec} \ll H(T_p)$,  the reheating process is supposed to work so slowly that $\chi$ is allowed to roll down and oscillate around the true vacuum until $\Gamma_{\rm dec} \sim H(T_p)$, 
where the universe undergoes the matter-dominated period, 
In that case, the reheating temperature $T_{r}$ reads~\cite{Hambye:2018qjv,Ellis:2019oqb,Kierkla:2022odc}
\begin{equation}
	T_{r}\simeq T_{\rm vac} \sqrt{\frac{\Gamma_{\rm dec}}{H(T_p)}}\,.
\label{Treh-Tvac}
\end{equation}

\end{itemize}

\begin{figure}[htbp]
  \centering
  \includegraphics[scale=0.56]{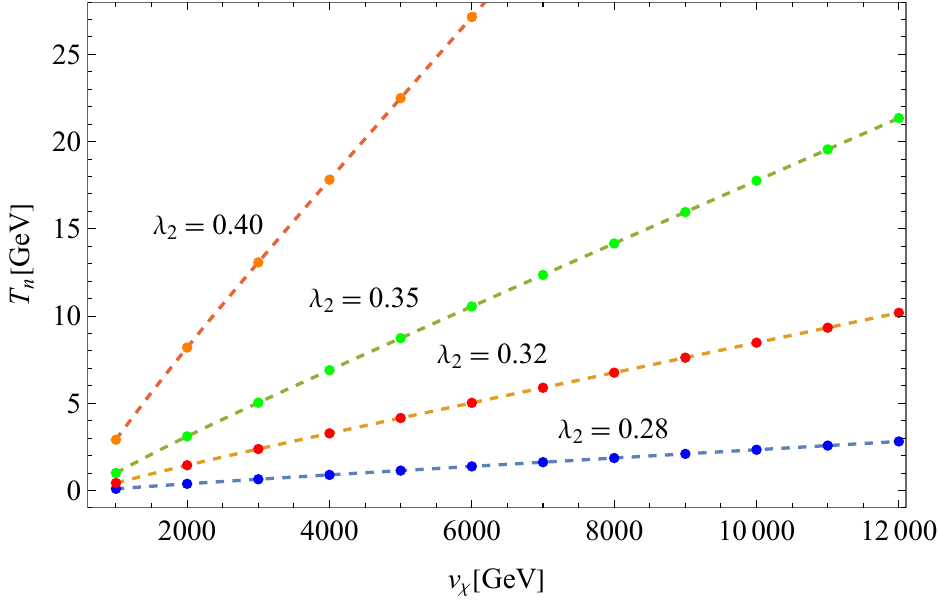}
  \includegraphics[scale=0.562]{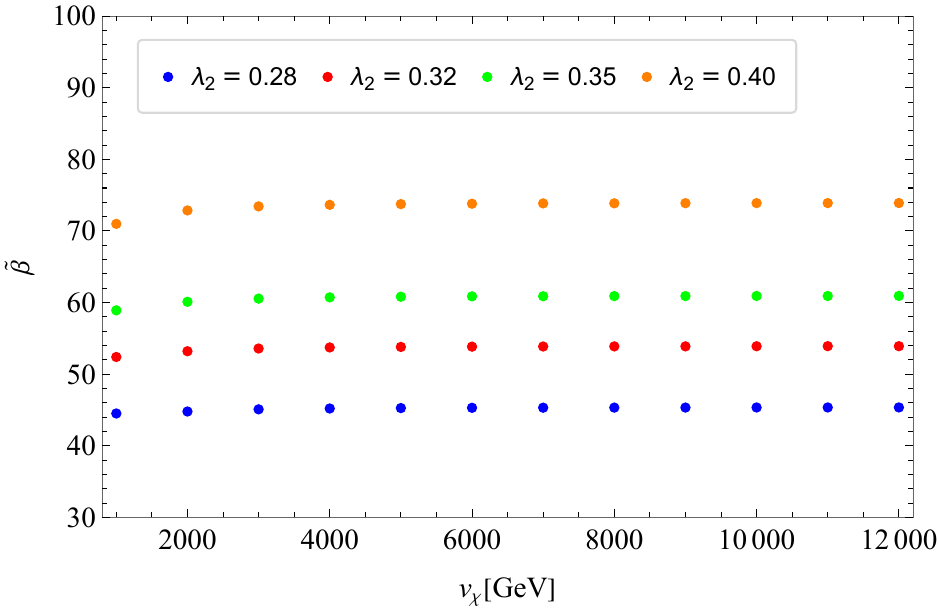}
  \caption{Left: The plot of $T_n$ vs. $v_\chi$ with $\lambda_2$ varied 
  in the allowed range as in Fig.~\ref{fig:nucleation}; 
  Right: $\tilde{\beta}$ vs. $v_\chi$ with the same varied range of $\lambda_2$. }
  \label{Tnvsvchi}
  
\end{figure}

As will be discussed in more details, however, in the present study we do not refer to  
the size of the $\chi$ decay rate in addressing the reheating process as classified 
in way as above. 
More crucial to notice is that at any rate whether the case is (i) or (ii), 
$T_{r}$ almost simply scales as (see also Eq.(\ref{V1}))
\begin{align}
   T_{r} \propto T_{\rm vac} \propto \lambda_2^{1/2} v_\chi 
    \,. \label{scaling:Treh}
\end{align}
Having this scaling law in our mind, 
we now discuss the correlation between the cosmological phase transition parameters, 
$\alpha$ and $\beta$, and the nucleation temperature $T_n$ or $T_p$. 
First of all, see two panels in Fig.~\ref{Tnvsvchi}. 
In the left panel the $v_\chi$ dependent on $T_n$ varying $\lambda_2$ is plotted within the allowed regime as in Fig.~\ref{fig:nucleation}, while the right panel shows the $v_\chi$ dependence on $\beta$ for fixed $\lambda_2$.  
In the left panel we observe that $T_n$ linearly grows with $v_\chi$ for any $\lambda_2$.  
This trend is closely tied with the scalegenesis feature~\footnote{  
To be phenomenologically realistic, the scale invariance must be approximate even at the 
classical level. 
However, the general trends addressed here will not significantly be affected 
as long as small enough explicit scale breaking is taken into account, as noted also in the previous section. See Summary and Discussion, for more details.}:  
only one dimensionful parameter $v_\chi$ is dominated after the dimensional transmutation, 
hence at finite temperature the dimensionless bubble action can be almost fully controlled by 
the dimensionless ratio $v_\chi/T_n$, once $\lambda_2$ is fixed,  which 
means $T_n$ linearly changes with the variation of $v_\chi$. 
This can more quantitatively be viewed as follows:  
given that $S_3/T_n$ is a function of $(v_\chi/T_n)$, which is fixed to $\simeq 140$, 
then the stationary condition of $S_3/T_n$ leads to $d v_\chi/d T_n = v_\chi/T_n$, 
hence $T_n \propto v_\chi$. 
Likewise, one can prove that $\tilde{\beta}$ is insensitive to increasing $v_\chi$ as plotted in 
the right panel of Fig.~\ref{Tnvsvchi}. This trend can be understood by noting 
that $\tilde{\beta} = T_n \partial (S_3/T_n)/\partial T_n = - v_\chi/T_n \partial (S_3/T_n)/\partial (v_\chi/T_n)$.

Second, we recall the scaling property of $T_{r}$ in Eq.(\ref{scaling:Treh}), 
$T_{r} \propto v_\chi$. 
Since both $T_n$ and $T_{r}$ linearly grow with $v_\chi$, $\alpha$ defined 
as in Eq.(\ref{alpha}) follows the same trend as what $\tilde{\beta}$ does. 
Thus we have 
\begin{align}
    \alpha \sim {\rm const.} \,, \qquad \tilde{\beta} \sim {\rm const.}\,, \qquad \textrm{in } v_\chi\,.  
\label{scalegenesis-trend}
\end{align}
Note, furthermore, that for a larger $\alpha$ as in Eq.(\ref{Treh-alpha}), 
the slope of $T_n$ with respect to $v_\chi$ is almost completely fixed as $\alpha^{-1/4}$. 
Those cosmological phase transition features are thus characteristic to the (almost) 
scale invariant setup.

In comparison, 
in the literature~\cite{Kawana:2022fum} with a similar scale-invariant setup, 
$\alpha$ and $\tilde{\beta}$ have been evaluated at not $T=T_n$, but at $T_p$, 
where the latter does not exhibit a simple scaling property with respect to $v_\chi$ 
unlike the former. 
Therefore, in the literature $\alpha$ and $\tilde{\beta}$ look sensitive to 
increase of $v_\chi$. 
The discrepancy between $T_n$ and $T_p$ is thought to become significant 
when the GW production  
is addressed with the reheating process taken into account. 
A conventional estimate will be based on the instantaneous reheating 
with $T_{r} \simeq (1 + \alpha)^{1/4} T_p$ as in Eq.(\ref{Treh-alpha}). 
Assuming the entropy conservation involving the reheating epoch 
one may then get the redshifted GW spectra and frequency at present day, 
which are scaled with $T_p$. 
However, as we will clarify more explicitly in the next section, 
it turns out that it is not $T_p$ or $T_n$ but $T_{r}$ that 
sets the scale of the GW spectra and frequencies.

\section{GW production: prospects for nano HZ and higher frequency signals}

In this section, we discuss the stochastic GW backgrounds sourced by the ultra-supercooling produced in the present model setup. 
The resultant GW spectrum ($\Omega_{\rm GW} h^2$) comes from three processes: the collisions of bubble walls 
($\Omega_{\rm coll} h^2$), the sound waves in the plasma 
($\Omega_{\rm sw} h^2$), 
and the magnetohydrodynamics turbulence in the plasma 
($\Omega_{\rm turb} h^2$), i.e., 
\begin{equation}
	\Omega_{\rm GW} h^2 = \Omega_{\rm coll} h^2 + \Omega_{\rm sw} h^2 + \Omega_{\rm turb} h^2\,,
\end{equation}
where $h$ is the Hubble constant in units of 100 km/(s$\cdot$Mpc).

In the ulta-supercooled phase transitions with $\alpha$$\gg 1$, the transition temperatures are low enough that the friction induced from the the plasma is too small to stop the bubble wall accelerating before it collides with other bubbles. Therefore, most of the released latent heat flows into the bubble walls and accelerates the bubbles without being bound, hence runs away~\cite{Bodeker:2009qy,Caprini:2015zlo} almost with 
the speed of light $v_w \sim c$. 
Thus we see that the bubble collisions give the dominant contribution to the GW spectrum. 
The efficiency factor $\kappa_{\rm coll}$, which characterizes the energy transfer between the vacuum energy and the kinetic energy of the bubble wall, reads
\begin{equation}
	\kappa_{\rm coll}=1-\frac{\alpha_\infty}{\alpha},\quad \alpha_\infty\simeq \frac{30}{24\pi^2}\frac{\sum_i c_i \Delta m_i^2}{g_p T_p^2}\,, 
\label{kappa}
\end{equation}
where the sum running over $i$ counts all relativistic particles in the false vacuum and all heavy and nonrelativistic ones in the true vacuum; $\Delta m_i^2$ is the difference of their (field-dependent) squared masses; $g_p$ corresponds to the effective d.o.f. for the relativistic particles in the false vacuum; $c_i$ is equal to $n_i$ as in Eq.(\ref{A-B}) for bosons and $\frac{1}{2}n_i$ for fermions, with $n_i$ being the number of the d.o.f. for species $i$. 
In the present model, which predicts the ultra-supercooling, 
we can safely take $\kappa_{\rm coll} \sim 1$, which is due to the fact that 
\begin{equation}
    \frac{\alpha_\infty}{\alpha}\propto \frac{T_p^2}{v_\chi^2}\ll 1\,.
\label{alpha-limit}
\end{equation}


We also need to take into account the redshift factor ($\frac{a_p}{a_0}$) which describes the Hubble evolution acting on 
the GWs from when it is produced at the epoch corresponding to 
the scale factor $a_p$ up until today at $a_0$. 
We intercept $\frac{a_p}{a_0}$ by the epoch ($a_r= a(T_{r})$), at which the latent heat released from the false vacuum starts to get efficient enough to be converted into the radiation, to be dominated over the universe (regarded as the end of the reheating): $\frac{a_p}{a_0} = \frac{a_p}{a_r} \cdot \frac{a_r}{a_0}$~\footnote{
For the exponential nucleation phase transitions as in the present model case, the percolation temperature $T_p$ should not be so much below the temperature at which bubbles collide. 
Therefore, it is appropriate to choose the temperature at which GWs are produced as the percolation temperature. 
}. 
Since the reheating process is nonadiabatic and cannot simply be 
described by thermodynamics, we instead of temperature monitor $\frac{a_p}{a_r}$ in terms of 
the e-folding number $N_e$, which is accumulated during the period from when one bubble is nucleated up to the end of the reheating~\footnote{
A similar evaluation of the redshift factor in terms of the e-folding number 
has been made in~\cite{An:2022toi}, which is applied to the inflationary epoch, not 
the reheating process that, instead, the authors assumed 
to be matter dominated or kination dominated. 
}. 
The latter part, $\frac{a_r}{a_0}$, is totally thermal, hence can simply be scaled by the entropy conservation per comoving volume: $s(T_{r})a_{r}^3=s(T_0)a_0^3$ with 
the thermal entropy density $s(T) = \frac{2\pi^2}{45} g_{*s}(T) T^3$.

One might think about constructing a couple of the Boltzmann equations with respect to 
the radiation energy density and the energy densities of $\chi$ and the SM Higgs, to which 
$\chi$ decays via the Higgs portal, 
and evaluate what is like ``matter-radiation" equality at which the reheating 
temperature $T_r$ can be defined. However, this approach cannot go beyond 
the level of the ensemble average approximation of the dynamics, i.e., sort of a classical 
level not incorporating the nonadiabatic and nonperturbative relaxation dynamics till 
the universe is fully radiated starting from the end of the supercooling in 
the de-Sitter expansion. 
Thus, there would be still lots of uncertainties involved if 
one addresses the reheating by naively referring to such Boltzmann equations with 
the size of the $\chi$ decay rate. 
Therefore, at this moment in our best reasonable way, 
we parametrize the epoch during the reheating process by the e-folding, as noted above, 
and simply assume the instantaneous reheating without referring to the size of 
the $\chi$ decay rate as classified in Eqs.(\ref{Treh-alpha}) and (\ref{Treh-Tvac}).

Thus at this moment we write the redshift factor as   
\begin{align}
    \frac{a_p}{a_0} = \frac{a_p}{a_r}\frac{a_r}{a_0}
	&= e^{-N_e} \cdot 
 \frac{g_0^{1/3} \cdot T_0}{g_r^{1/3} \cdot T_r}  
 \,, \label{new-formula}
\end{align} 
where $g_0\simeq 2+ \frac{4}{11}\times \frac{7}{8}\times 2N_{\rm eff}$ with $N_{\rm eff}=3.046$~\cite{ParticleDataGroup:2022pth} and $g_{r}$ are the d.o.f. at the present-day temperature $T_0=2.725{\rm K}$ and at the reheating temperature, respectively. 
The effective d.o.f. for the entropy density and of energy density has been assumed to be identical 
each other, i.e., assuming no extra entropy production other than the one created passing 
through the reheating epoch. 

To make comparison with the conventional formula of the peak frequency, based on inclusion of 
the entropy conservation during the reheating epoch~\cite{Caprini:2015zlo}, 
\begin{equation}
	f_{\rm coll} \Bigg|_{\rm conventional} =   1.65\times 10^{-5}{\rm Hz}\times\left(\frac{0.62}{v_w^2-0.1v_w+1.8}\right)\times \left(\frac{\beta}{H(T_p)}\right)\left(\frac{T_p}{100\,{\rm GeV}}\right)\left(\frac{g_p}{100}\right)^{\frac{1}{6}}\,,  
\label{conventional}
\end{equation}
we rewrite Eq.(\ref{new-formula}) as follows: 
\begin{align}
	\frac{a_p}{a_0} = \frac{a_p}{a_r}\frac{a_r}{a_0}
	&= e^{-N_e}\frac{g_0^{1/3}T_0}{g_r^{1/3} T_r}H(T_r)\frac{H(T_p)}{H(T_r)}\frac{1}{H(T_p)} \notag\\
	&= e^{-N_e}\frac{g_0^{1/3}T_0}{g_r^{1/3} T_r}\frac{g_r^{1/2}\pi T_r^2}{3\sqrt{10}M_{\rm pl}}\frac{H(T_p)}{H(T_r)}\frac{1}{H(T_p)} \notag \\
	&= e^{-N_e}\left(\frac{\rho(T_p)}{\rho(T_r)}\right)^{1/2}\frac{100^{7/6} g_0^{1/3}\pi T_0}{ 3\sqrt{10}M_{\rm pl}}\left(\frac{g_r}{100}\right)^{1/6}\frac{T_{r}}{100{\rm GeV}}\frac{1}{H(T_p)}\,, 
\end{align}
where we have used 
the Friedmann equations $3M_{\rm pl}^2 H_{r}^2=\rho(T_{r})=\frac{\pi^2}{30}g_{r}T_{r}^4$ and $3M_{\rm pl}^2 H_p^2=\rho(T_p)$. 
The redshifted peak frequency is thus evaluated as
\begin{equation}
	f_{\rm coll} = e^{-N_e}\left(\frac{\rho_p}{\rho_r}\right)^{1/2}\times 1.65\times 10^{-5}{\rm Hz}\times\left(\frac{0.62}{v_w^2-0.1v_w+1.8}\right)\times \left(\frac{\beta}{H(T_p)}\right)\left(\frac{T_{r}}{100\,{\rm GeV}}\right)\left(\frac{g_p}{100}\right)^{\frac{1}{6}}\,, 
\end{equation}
which is compared to the conventional formula in Eq.(\ref{conventional}): 
\begin{align}
  f_{\rm coll} = e^{-N_e}\left(\frac{\rho_p}{\rho_r}\right)^{1/2} \left( \frac{T_{r}}{T_p} \right) \times f_{\rm coll} \Bigg|_{\rm conventional}  
  \,. \label{f-formula}
\end{align}

This implies that even when the GW is produced at the QCD scale or so, 
the nano Hz frequency is unlikely to be realized.  
One might still suspect  that if an inflationary stage, after the tunneling for the flat enough Coleman-Weinberg type potential, is present, 
it could suppress the peak frequency due to a huge amount of the accumulated e-folding $N_e$, so that the nano Hz signal could be generated. 
However, this would not be the case: the tunneling exit point is supposed to be within the inflation region, which requires that the coupling $\lambda_2$ is tiny enough that no percolation takes place and the stationary condition $B>0$ in Eq.(\ref{A-B}) is also violated, thus no bubble collision, nor GWs induced from the first-order phase transition. 
Thus it is clarified that the peak frequency is shifted to higher by scaling with $T_{r}$ (that is, 
``blueshifted").

GW spectra sourced from the bubble wall are evaluated based on the 
simulations of bubble wall, leading to the following fitting function with the conventional redshift incorporated~\cite{Huber:2008hg}:
\begin{equation}
	\Omega_{\rm coll} h^2 \Bigg|_{\rm conventional}=  
 1.67\times 10^{-5} \left(\frac{H(T_p)}{\beta}\right)^2\left(\frac{\kappa_{\rm coll}\,\alpha}{1+\alpha}\right)^2\left(\frac{100}{g_p}\right)^\frac{1}{3}\times \left(\frac{0.11v_w^3}{0.42+v_w^2}\right)S_{\rm coll}(f)\,,
\end{equation}
where $S_{\rm coll}(f)$ parametrizes the spectral shape, which is given also by the fitting procedure to be~\cite{Huber:2008hg}
\begin{equation}
	S_{\rm coll}(f) = \frac{3.8\left(\frac{f}{f_{\rm coll}}\right)^{2.8}}{1+2.8\left(\frac{f}{f_{\rm coll}}\right)^{3.8}}\,.
\end{equation}
These GWs also get redshifted similarly to the peak frequency as 
\begin{align}
    \Omega_{\rm coll} h^2  =  e^{-4N_e} \left( \frac{\rho_p}{\rho_r} \right) \times \Omega_{\rm coll} h^2 \Bigg|_{\rm conventional}
\,, \label{GW-formula}
\end{align}
which generically tends to get suppressed by the e-folding $N_e$ and $(\rho_p/\rho_r)$.

\begin{figure}[t]
	\includegraphics[scale=0.9]{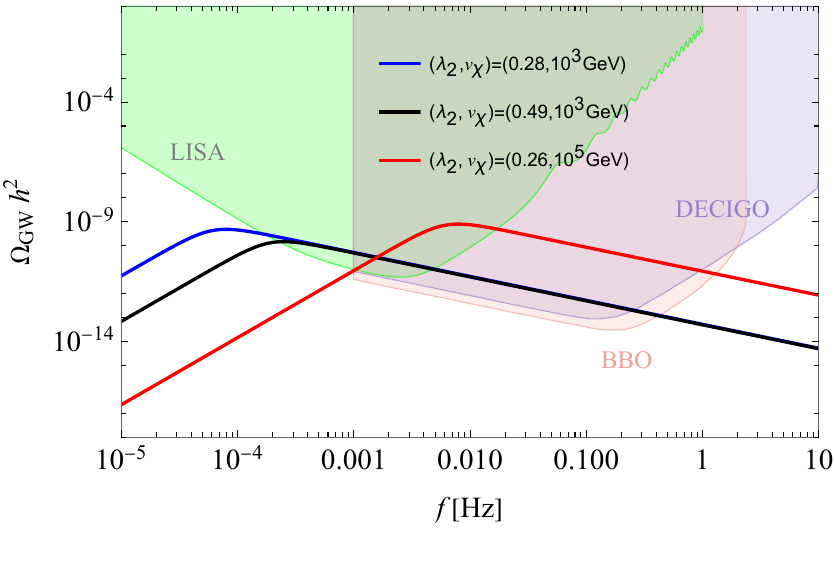}
	\caption{\label{fig:GWspectra} 
		The plot on GW power spectra for several benchmarks of the present model with $N_f =8$ 
  in comparison with future prospected detector sensitivities~\cite{Moore:2014lga,Schmitz:2020syl}. 
  } 
\end{figure}

From the refined formulae Eqs.(\ref{f-formula}) and (\ref{GW-formula}), 
we see that $f_{\rm coll}$ linearly grows as $v_\chi$ because 
$T_{r}$ gets larger as $v_\chi$ gets larger as in Eq.(\ref{scaling:Treh}), while $\Omega_{\rm GW} h^2$ is insensitive to increase of $v_\chi$. 
In Fig.~\ref{fig:GWspectra} we plot the GW spectra for several values of 
$(\lambda_2, v_\chi)$ for $N_f=8$ 
with the instantaneous reheating ($\rho_p = \rho_r$ and $N_e=0$) assumed 
in Eqs.(\ref{f-formula}) and (\ref{GW-formula}). 
together with the prospected sensitivity curves~\cite{Moore:2014lga,Schmitz:2020syl}. 
As evident from the newly proposed formula on the peak frequency in Eq.(\ref{f-formula}), 
the GW peaks are generically shifted toward higher due to the significant 
dependence of $T_{r} (\propto v_\chi)$, in comparison with 
a similar scalegenesis prediction in the literature~\cite{Kawana:2022fum}.
In fact, the displayed three GW signals have been sourced from 
the ultra-supercooled first-order phase transitions at 
lower nucleation/percolation temperatures $T_p=100$ MeV (for blue curve) and 10 GeV (for both black and red curves), 
which are typically thought to be low enough to realize the GW peak signals around nano Hz simply 
following the conventional formula in Eq.(\ref{conventional}). 
Nevertheless, the produced signals following the proposed formula Eq.(\ref{f-formula})
peak at much higher frequencies, say, ranged from $10^{-4}$ Hz to $10^{-2}$ Hz, as seen from 
Fig.~\ref{fig:GWspectra}. 
This is manifested by the linear $T_{r}$ dependence in the peak frequency formula, Eq.(\ref{f-formula}), in which currently we have $T_{r} \simeq 41$ GeV (for blue curve), 70 GeV (for black curve), and 5.2 TeV (for red curve), respectively. 
Consequently, even the smaller $v_\chi$ (i.e. lower new physics scale $\sim 1$ TeV) 
can easily reach the LISA prospect and other higher frequency prospects (BBO and DECIGO, and so forth), 
though the GW signals would generically be as small as the lower bounds of the prospects.

On the other side of the same coin, we can conclude that 
nano (or less nano) Hz signals cannot be reached by 
the ultra-supercooled scalegenesis of this sort, because of the inevitable ``blueshift" of 
the GW frequency: if the nano HZ signal is imposed to realize, i.e., simply $T_{r} \sim 100$ MeV, then Eq.(\ref{Treh-Tvac}) requires $v_\chi \sim 1$ TeV with $g_{r} = {\cal O}(100)$, which leads to 
$\alpha \gg 1$, hence extremely tiny $T_n$ or $T_p$. 
Thus $T_p$ would be required to be around $\sim$ MeV or less, 
which is actually inside the excluded regime with no percolation (See Fig.~\ref{fig:nucleation}). 


The large $N_f$ models, e.g., with $N_f=8$, as what we currently focus on, 
tends to make $\tilde{\beta}$ larger, while 
the $N_f$ dependence in $\alpha$ gets almost insensitive in the GW signals sourced from 
collisions, because anyhow the ultra-supercooling merely provides huge $\alpha$ as noted 
around Eqs.(\ref{kappa}) and (\ref{alpha-limit}) to give $\kappa_{\rm coll} \sim 1$ 
irrespective to the precise large number of $N_f$. 
Thus, the large $N_f$ case tends to further ``blueshift" 
the peak frequency of the GW and make the GW signal strength smaller, 
due to the produced large $\tilde{\beta}$.

\section{Summary and discussion}

In this paper, we have discussed GW spectra predicted from 
the electroweak scalegenesis of the Higgs portal type, what we call the Higgs portal scalegenesis, 
and payed a particular attention into a dark sector having a large number of dark chiral flavors. 
We have modeled the dark sector by the linear sigma model description, 
which has the chiral $U(N_f)_L \times U(N_f)_R$ symmetry and is 
coupled to the SM Higgs via the Higgs portal 
with keeping the classical scale invariance. 
Working on the Gildener-Weinberg mechanism, 
we have observed that models of this class undergo 
a strong first-order chiral phase transition and 
ultra-supercooling. 
Evaluation of the bubble nucleation/percolation has clarified 
the possibility of generation of GWs sourced from the ultra-supercooling, 
which is accessible in a wide parameter space of the model (see Fig~\ref{fig:nucleation}). 
We also clarified the characteristic features for $\alpha$ and $\tilde{\beta}$ stemming from 
the consequence of the Higgs-portal scalegenesis irrespective to the case with or without 
large dark-sector flavors (Eq.(\ref{scalegenesis-trend})).

Our particular emphasis has also been provided in evaluation of the reheating and relaxation processes, 
which necessarily and significantly arise from the ultra-supercooling predicted 
from the scale-invariant setup of the present class of models. 
Though such a reheating epoch is generically non-thermal and nonadiabatic, we proposed to parametrize 
it in terms of the e-folding number, not assuming 
the conventional entropy conservation as in thermodynamic cosmology. 
This approach has derived refined formulae for the redshift effects on 
the peak frequencies of the produced GWs and the GW signal strengths 
(Eqs. (\ref{f-formula}) and (\ref{GW-formula})). 
Particularly, it has been clarified that the peak frequencies finally get ``blueshifted", 
in comparison with the conventional approach based on the thermal entropy conservation 
including the reheating epoch (see Eq.(\ref{f-formula})).

We then observed that the ultra-supercooling predicted from the Higgs-portal scalegenesis   
yields none of nano Hz GW signal, instead, prefers to give the higher frequency signals (Fig.~\ref{fig:GWspectra}). 
This is irrespective to whether $N_f$ is large or small, that is in a sense rather generic conclusion simply 
because the model is the ultra-supercooling and the Higgs portal: 
the ultra-supercooling generically has the trend in such a way that 
a smaller $v_\chi$ could get smaller $T_n$, 
hence a small enough peak frequency``blueshifted" by $T_r$ to achieve nano Hz. 
However, the smaller $v_\chi (\lesssim 1\,{\rm TeV})$ has already been excluded by the Higgs coupling measurement 
(Eq.(\ref{Higgs-coupling})), 
so that the peak frequencies inevitably have to be higher. On the other side, $T_n$ cannot be small with keeping 
large enough $v_\chi$, otherwise one gets null percolation (Fig~\ref{fig:nucleation}).  
It is interesting to note that indeed, all so far proposed beyond the SMs with successful nano Hz GWs do not possess ultra-supercooling. 
One way to escape from this dilemma may be to cease reheating released back to the universe by making the latent heat almost transported into another sector. Pursuing this type of way out deserves another publication.

It seems to be robust that no nano Hz signals are predicted at this point, but 
the ultra-supercooling Higgs-portal scalegenesis still keeps the future prospected detection sensitivity, like at LISA, BBO, and DECIGO, etc. 
We also found that with large chiral flavors in the dark sector, the GW signals are made further smaller 
and the peak frequencies higher.

In closing, we give comments on the explicit chiral and scale breaking, 
which is, to be phenomenologically viable, necessary to be incorporated into the present model. 
The explicit breaking needs to be so small that the flat direction we have worked on 
is still approximately operative in searching for the true vacuum, 
as has been addressed in the literature~\cite{Iso:2014gka,Ishida:2019wkd,Zhang:2023acu,Cacciapaglia:2023kat}. 
Given such a small enough breaking term, the model 
will predict light pseudo Nambu-Goldstone bosons (dark pions) with  
the number of $N_f^2$ or $N_f^2-1$ (with one decoupled due to the axial anomaly), 
depending on the 
underlying theory for the linear sigma model description. 
The dark-pion mass term plays a role of 
the tadpole term for the $\chi$ potential does, 
which will make the false vacuum shifted depending on 
temperature until the supercooling ends, as was clarified in~\cite{Zhang:2023acu}. 
The origin of such a chiral and scale breaking could be linked with 
presence of a gravitational dilaton which couples to the underlying 
dark QCD fermion bilinear $\bar{F} F$: 
$V_{\rm tadpole} = {\cal C} \cdot e^{\varphi} \bar{F}F \approx 
- {\cal C} \cdot e^{\varphi} \langle - \bar{F}F \rangle
\cdot {\rm Tr}[M^\dag + M] =
- {\cal C}\cdot \langle - \bar{F}F \rangle \cdot \chi \cos\theta + \cdots $ with $\varphi$ and ${\cal C}$ being the dilaton and a constant coupling, respectively. 
As long as the size of the $\chi$-quartic coupling $\lambda_2$ is sizable enough as in 
the desired regime displayed in Fig.~\ref{fig:nucleation}, 
both the percolation and nucleation can be realized not substantially to  
affect what we have addressed and clarified so far in the scale-invariant limit. 

The precise size of the dark pion mass ($m_{\pi_d}$) highly depends on the underlying 
theory. Generically the dark pions will be stable to be a dark matter candidate 
if and only if the dark isospin symmetry is not violated. The dark pions ($\pi_d^A \equiv (\eta, \pi^a)$ in Eq.(\ref{matrixfield})) are expected to 
be light and can couple to the SM particles via the Higgs portal with $\lambda_{\rm mix}$, which is 
$\lesssim 10^{-3}$ for $v_\chi \gtrsim 1$ TeV (See Eq.(\ref{lambda-mix})). 
Therefore the dark pions can be pair-produced at the LHC via the Higgs production processes like 
$pp \to h  \to \chi^* \to \pi_d^A \pi_d^A$ with the final state identified as a large missing energy.  
The cross section is fixed by the size of $v_\chi$, 
$\lambda_2$, $m_{\pi_d}$, and $N_f$. 
Hunting the pseudo-dilaton $\chi$ is also accessible at the LHC via the Higgs portal 
coupling. 
Those would be interesting studies in light of the LHC-run 3 with high luminosity, 
hence would provide a complementary probe of the Higgs-portal scalegenesis with large flavors including 
the collider constraints on the dark pion mass, 
which is to be explored elsewhere.

In the thermal history, the dark pions as dark matters can be produced 
via the annihilation into the lighter SM particles, presumably, diphoton, via the Higgs 
portal coupling including the $\chi$ exchange: $\pi_d^A \pi_d^A \to \chi^* - h^* \to \gamma\gamma$. 
Still, this process needs to assume the portal coupling to be thermalized with 
the SM thermal plasma, which can be evaluated by equating the $\chi \to h$ conversion rate $\Gamma_{\chi \to hh}$ 
and the Hubble rate. The conversion rate is roughly estimated as 
$\Gamma_{\chi \to h} = n_\chi \langle \sigma v \rangle 
\sim \frac{\lambda_{\rm mix}^2}{m_\chi^2} (m_\chi T)^{3/2} e^{- m_\chi/T}$. 
We take 
$m_\chi = \sqrt{\frac{\lambda_2}{2 \pi^2} } v_\chi \gtrsim 1.2\,{\rm TeV}\times (\frac{\lambda_2}{0.3})^{1/2}$ (see Eq.(\ref{Mchi}) and Figs.~\ref{fig:nucleation} and \ref{fig:GWspectra} for the reference value of $\lambda_2$) and 
$|\lambda_{\rm mix}| = m_h^2/v_\chi^2 \lesssim 10^{-3}$ for $v_\chi \gtrsim 1$ TeV. Comparing this conversion rate with the radiation-dominated Hubble rate 
$H \sim \sqrt{g_*(T)}\cdot T^2/M_{\rm pl}$ with $g_*(T)\sim 100$, 
we see that 
the conversion is thermally decoupled at $T \sim 50$ GeV for $v_\chi \sim 1$ TeV. 
As $v_\chi$ gets larger, the decoupling temperature will be higher. 
Compared to the thermal freeze-out of the dark pion annihilation 
which is expected to happen usually when $m_{\pi_d}/T \sim 20$, 
it turns out that the dark pion annihilation will be stopped at higher $T$ 
before the conventional annihilation process is frozen out. 
This implies that the dark pions keep the thermal number density  
and decouples from the SM plasma at $T = T_{\rm dec}> 50$ GeV, and adiabatically diluted to reach today 
following the entropy conservation in the dark thermal plasma with ${\cal T}$ isolated from the SM thermal plasma with $T$. 
From the entropy conservation for each plasma cooled down from $T={\cal T}=T_{\rm dec}$, 
we find the dark-sector temperature always gets much lower as 
$({\cal T}/T)^3 = g_{* s}^{\rm SM}(T)/g^{\rm SM}_{*s}(T_{\rm dec}) $.  
The dark sector keeps the isolated own thermal plasma until today,   
so the dark pion yield at present is highly suppressed to be negligibly small.

Since the SM Higgs is a nonrelativistic particle at $T < 50$ GeV, there would be no chance to make the freeze-in mechanism~\cite{Blennow:2013jba,Hall:2009bx} work for the dark pion as well. 
Thus the thermal relic abundance is unlikely efficiently produced, hence the thermal dark pions are not expected to explain 
the dark matter density today.

The dark pions could still be produced non-thermally via the coherent oscillation mechanism, 
just like axionlike particles. 
Since the dark pions can develop the potential of the cosine form, $V(\pi_d) = m_{\pi_d}^2 f_{\pi_d}^2 
(1- \cos \frac{\pi_d}{f_{\pi_d}})$, where, perhaps, $\ll v_\chi$ for many flavor QCD case~\cite{Ishida:2019wkd,Zhang:2023acu}. 
The size of the energy density per flavor accumulated by the coherent oscillation 
depends on $f_{\pi_d}$, $m_{\pi_d}$, and the initial place of the dark pion (so-called the misalignment angle), which would be highly subject to the modeling of the underlying 
theory, to be pursued in another publication.

\section*{Acknowledgments} 

This work was supported in part by the National Science Foundation of China (NSFC) under Grant No.11747308, 11975108, 12047569, 
and the Seeds Funding of Jilin University (S.M.), 
and Toyama First Bank, Ltd (H.I.).

\end{document}